\documentclass[showpacs,prb,aps,twocolumn]{revtex4}
\usepackage{graphicx}
\usepackage{color}
\usepackage{dcolumn}
\usepackage{bm}
\def\ph2{{\it p}-H$_2$}
\def\n{\mathbf{n}}
\def\np{\mathbf{n'}}
\def\gapx{\lower 2pt \hbox{$\buildrel>\over{\scriptstyle{\sim}}$\ }}
\def\lapx{\lower 2pt \hbox{$\buildrel<\over{\scriptstyle{\sim}}$\ }}
\begin{document}
\title{Ground-State Properties of Quantum Many-Body Systems: Entangled-Plaquette States and Variational Monte Carlo}

\author{Fabio Mezzacapo$^1$}
\author{Norbert Schuch$^1$}
\author{Massimo Boninsegni$^2$}
\author{J. Ignacio Cirac$^1$}

\affiliation{$^1$Max-Planck-Institut f\"ur Quantenoptik, Hans-Kopfermann-Str.1, D-85748, Garching, Germany\\
$^2$Department of Physics, University of Alberta, Edmonton, Alberta, Canada T6G 2G7 }
\date{\today}

\begin{abstract}
We propose a new  ansatz for the ground-state wave function of quantum many-body systems on a lattice. The key idea is to cover the lattice with plaquettes and obtain a  state whose
configurational weights can be optimized by means of a Variational Monte Carlo algorithm. Such a scheme applies to any dimension, without any ``sign" instability. We show results for various  two dimensional spin models (including frustrated ones).  A detailed comparison with available exact results, as well as with  variational methods based on different ansatzs is offered. In particular, our numerical estimates are  in quite good agreement with exact ones for unfrustrated systems, and compare favorably to other methods for frustrated ones. 
\end{abstract}


\maketitle

\section{Introduction} 

The study of the ground-state (GS) properties of quantum many-body systems is one of the most challenging tasks of theoretical physics. Exact results can be obtained numerically only for systems of relatively  small size (i.e., few particles); this limitation is particularly severe, e.g., when studying phase transitions, wherein the emergence of long-range order can only be established by carrying out an extrapolation of the estimates to the thermodynamic limit.
Two main numerical techniques, namely  Density Matrix Renormalization Group (DMRG)\cite{dmrg} and Quantum Monte Carlo (QMC)\cite{qmc} have successfully been employed to investigate quantum spin models on a lattice.  These computational approaches, however, find their optimal applicability under different specific constraints. DMRG yields extremely accurate results in one dimension (1D) even for very large systems,  but fails in describing the properties of the quantum GS in higher dimension due to  the unfavorable scaling with the system size of the computational resources needed.\cite{liang}  QMC, on the other hand, is the method of choice for quantum systems obeying Bose statistics in any dimension, but suffers from the notorious ``sign problem" for Fermi systems.

Generalizations of the variational family of Matrix-Product States (MPS) underlying the DMRG have recently been investigated to go beyond the discussed limitations of DMRG itself and QMC. Specifically the most natural extension of MPS is given by Projected-Entangled Pair States (PEPS)\cite{peps} which efficiently approximate ground states of local Hamiltonians\cite{has} and have been used to simulate, otherwise intractable, 2D quantum systems.\cite{peps1,orusetal} Other variational families of states have also been proposed and tested in 2D.\cite{vidalmera, Levin, Weng, LevinWen, cincio, vidal2, vidl} Despite the promising results obtained so far, it seems very hard to use those methods in 3D [or even for systems with periodic boundary conditions (PBC)] due to the unfavorable scaling of the required computer resources.

A new possibility has recently emerged to combine the main advantages of DMRG and Monte Carlo in order to build new algorithms.\cite{SBS, SV} Some of them\cite{SBS, Snew} can be used in 2D and it seems that the one based on String-Bond States\cite{SBS} may be used even for some 3D systems.\cite{Alessandro} 

In this paper we introduce a new numerical technique that combines the strengths of QMC with an extension of PEPS to simulate lattices models, overcoming some of their limitations. We also test this technique with non-trivial models, and compare it with other techniques, including PEPS. Specifically, we propose a new class of states called Entangled-Plaquette States (EPS) which allow an accurate characterization of  the GS  of quantum spin systems by means of a simple Variational Monte Carlo (VMC) algorithm (see Ref. \onlinecite{VMC} for a general review).

The basic idea underlying  EPS can be schematically described as follows: Assume to cut a lattice in several sub-blocks (e.g. small PEPS) and extract for each sub-block the GS wave function. The wave function of the original system, expressed as the  product of the sub-block wave functions yields  reasonable (up to corrections which scale with the sub-block boundaries) estimates of GS energy and short-range (of the order of the sub-block size) correlations. These estimates dramatically improve if the sub-block size is increased and, more importantly, if  overlapping sub-blocks (i.e., entangled plaquettes) are employed.  The latter is the crucial point which allows, accounting for its correlated nature, a description of the quantum GS much more accurate than that obtainable with a simple non-overlapping-plaquette product state (i.e., in a mean-field fashion).\cite{o1} A GS wave function whose weights are   the product of variational parameters in one to one correspondence to the spin configuration of each entangled  plaquette,  appears,  consequently, the natural choice for a variational ansatz. Therefore, our numerical approach is based on a family of states, namely EPS, which share many analogies with PEPS, and takes advantage of Monte Carlo sampling to estimate physical observables of interest. It can be applied to  systems of any spatial dimensionality and, as a pure variational method (i.e., not involving imaginary time projection), is sign problem free.

We test  our numerical protocol on a variety of quantum spin  models on a square lattice  comprising as many as $N=L\times L = 400$ sites. Our energy estimates are in excellent agreement with those (``exact'' in practice) obtained by QMC for a system of lattice hard core bosons. In the presence of nearest-neighbor repulsion at the Heisenberg point, we find an extrapolated (to infinite lattice size)  value of the energy per site
which  differs from the QMC result\cite{sand} by less than $2\times$10$^{-3}$, and is  more accurate than VMC estimates obtained with a Jastrow wave function.\cite{cep90}
In the case of a frustrated antiferromagnet (i.e., the so-called $J_1-J_2$ model), for which a sign problem exists in QMC, our energy estimates (whose error relative to the exact ones is less than $1.5\times$10$^{-2}$) compare favorably with those obtained with PEPS, or fixed-node Green Function Monte Carlo (GFMC).\cite{massimo}

\section{Methodology}

Consider a collection of $N$ spins $\frac{1}{2}$ arranged on a $L\times L$ square lattice with $N=L^2$ (without loss of generality, here and in the following we will refer to this specific case). Provided a trial state $|\psi\rangle=\sum_{\mathbf{n}}W(\mathbf{n})|\mathbf{n } \rangle$ where $|\mathbf{n} \rangle = |n_1,n_2, \ldots , n_N\rangle$ and $n_i =  \pm 1$ $ \forall$  $i=1,\ldots,  N$, the energy expectation value on the given  state is:

\begin {equation}
\langle E \rangle=\frac{\sum_\mathbf{n}W^2(\mathbf{n})E(\n)}{\sum_\mathbf{n}W^2(\mathbf{n})}=\sum_\n P(\n)E(\n)
\label{energy}
\end {equation}
where
\begin{equation}
E(\n)=\sum_{\np} \frac{W(\np)}{W(\n)} \langle \np |H|\n \rangle, \mbox{    } P(\n)=\frac{W^2(\n)}{\sum_\n W^2(\n)}
\label{eq:2}
\end{equation}
and    $W(\n)= W^*(\n)$ (real weights are assumed for simplicity). According to the variational principle, $\langle E \rangle$ is an upper bound of the GS energy which can be evaluated by minimizing Eq. \ref{energy} with respect to the weights. At this point we have to make an ansatz for the wave function:  imagine to cover the lattice with plaquettes (say one of dimension $l_1\times l_2$ for each site)  and assign a coefficient $C_P^{\n_P}$ to all the possible $2^{l_1\times l_2}$ spin configurations of any single plaquette. Given a global spin configuration $|\n\rangle$, its weight can be expressed as follows:

\begin{equation}
W(\n)=\langle \n | \psi \rangle=\prod_{P=1}^N C_P^{\n_P}
\label{coeff}
\end{equation}
where $C_P^{\n_{P}}$ depends only on the spin state of the $\n_{P}$ sites belonging to the $P_{th}$ plaquette.
With this choice the analytic expression of the derivative of Eq. \ref{energy} with respect to $C_P^{\n_P}$ is:

\begin{equation}
\frac{\partial \langle E \rangle}{\partial C_P^{\n_P}}\!\!=\!2\!\sum_\n P(\n) \frac{1}{W(\n)}\frac{\partial W(\n)}{\partial C_P^{\n_P}}\Big[E(\n)-\sum_\np P(\np)E(\np)\Big].
\label{der}
\end{equation}

 The multidimensional summations in Eq. \ref{energy} and \ref{der} can exactly be evaluated only  for small $N$. For large systems ($N \gtrsim 30$) , one has to employ the Monte Carlo method. Specifically, the energy, as well as its derivatives, can be estimated from the same sample. Moreover, the only quantity depending on the plaquette coefficient with respect to which the derivative is taken is $D_P(\n_P)=\frac{1}{W(\n)}\frac{\partial W(\n)}{\partial C_P^{\n_P}}$. By using Eq. \ref{coeff} it turns out that $D_P(\n_P)$ is simply equal to $1/C_P^{\n_P}$ (easy and fast to compute).  Similarly to other works,\cite{SBS, SV}, the  steps of the basic variational algorithm adopted to calculate the GS energy are: i) Start from a randomly chosen initial configuration; ii) generate a large set of new configurations by flipping one or more spins via the Metropolis algorithm;\cite{metro} iii) evaluate the energy and its gradient vector; iv) update all the $C^{\n_P}_{P}s$ of a small step against the gradient direction; v) iterate from ii) until convergence of the energy is reached.  It is worth mentioning that expectation values of physical observables other than the energy can be evaluated according to Eq. \ref{energy} and \ref{eq:2} when $H$ is replaced by a generic operator $O$.

 For a single-spin flip the acceptance probability is given by:
  \begin{equation}
 A=\frac{\prod_{i}\Big [ C_i^{n_{1,i}^{old},n_{2,i}^{old},_{\cdots},n_{j,i}^{new},_{\cdots}, n_{l_1 \times l_2,i}^{old}}\Big ]^2} {\prod_i \Big [ C_i^{n_{1,i}^{old},n_{2,i}^{old},_{\cdots},n_{j,i}^{old},_{\cdots}, n_{l_1 \times l_2,i}^{old}} \Big]^2}
 \label{acc}
 \end{equation}
 where the flip is proposed for the $j_{th}$ spin and the products run over all the plaquettes which include such a spin.

\begin{figure}[h]
\centerline{\includegraphics[scale=0.6]{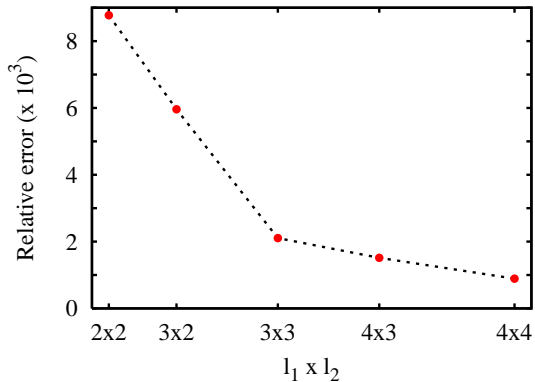}}
\caption{(color online)  Dependence on the plaquettes size of the error in the GS energy (computed with  the method illustrated in this work) relative to the  GFMC result for a system of hard core bosons at half filling on a $10\times10$ square lattice. PBC are assumed. The dashed line is only a guide to the eye. }
\label{fig:relerr}
\end{figure}

In a typical calculation, we start with $2\times2$ plaquettes; once the energy has converged,
the size of the plaquettes is increased to improve the estimate. The number of coefficients which need  be stored in memory for a spin-1/2 system is $N\times 2^{l1\times l2}$, and can be reduced taking into account problem-dependent symmetries. For each optimization step, a few thousands updates are necessary to get a rough estimate of the energy and efficiently move the coefficients along the gradient. At the later stages of the simulation, to reach the optimal energy value, an important role is played by the gradient step which has to be carefully tuned.
  An example of how the  error of the GS energy relative to the GFMC result decreases as a function of the plaquettes size is illustrated in Fig. \ref{fig:relerr}.

The relative error is already small (less than $1\%$) for $2\times 2$ plaquettes and reaches a value $ <10^{-3}$ when $4\times4$ plaquettes are used. Numerical data refer to a system of lattice bosons (at half filling) which interact via an infinite on-site (hard core) repulsion  on a $10\times10$ lattice with PBC.
Since the total magnetization along $z$ is a good quantum number, we performed the calculation in the canonical ensemble (i.e., in the $S^z=0$ sector). Consequently we chose to update the configuration by flipping pairs of spins $j$ and $k$ for which $S_j^z =-S_k^z$.
The expression of the acceptance probability for the pair update is a straightforward generalization of   Eq. \ref{acc}.

\section{Results}

Estimates of the GS energy per site  of a system of lattice hard core bosons,
for three different lattice sizes, are presented and compared to GFMC results in Tab. \ref{tab:hcb}.
Although the GFMC method is not purely variational (i.e., the GS wave function is projected out from  a trial state via imaginary time evolution) our numerical data are in excellent agreement with GFMC ones, even for the largest system  considered in this work ($L=20$).
\begin{table}[h]
\caption{\label{tab:hcb} GS energy per site (in units of the nearest-neighbor hopping integral $t$) for a system of hard core bosons
on a $L\times L$ lattice with PBC. For $L=20$,  $4\times3$ plaquettes have been employed, $4 \times4$ ones in the remaining cases. GFMC estimates have been also computed and are shown for comparison.}
\begin{ruledtabular}
\begin{tabular}{lcr}
$L$ & This work & GFMC \\
\hline
8& -1.1004(4)& -1.1008(3) \\
10 &-1.0988(4)  & -1.0998(1) \\
20 & -1.0952(7) & -1.0967(4) \\
\end{tabular}
\end{ruledtabular}
\end{table}

As a second case study we investigate the $J_1-J_2$ quantum spin Hamiltonian:
\begin{equation}
H= J_1\sum_{<i,j>}\mathbf{S}_i \cdot \mathbf{S}_j + J_2 \sum_{<<i,j>>}\mathbf{S}_i \cdot \mathbf{S}_j
\label{eq:heis}
\end{equation}
where the first (second) summation runs over nearest- (next-nearest-) neighbor sites. For $J_1=1$ and $J_2 = 0$ the above Hamiltonian describes the antiferromagnetic Heisenberg model, which is isomorphic to a system of hard core bosons with nearest-neighbor repulsion of strength $V$=$2t$. This model has been extensively studied numerically in the past (see for instance Refs. \onlinecite{sand} and \onlinecite{cep90}). GS energies per site for various lattice sizes are presented in Tab \ref{tab:heis} and compared to exact calculations\cite{exact}  and Stocastic Series Expansion (SSE) data.\cite{sand}
 \begin{table}[h]
\caption{\label{tab:heis} GS energy per site of the antiferromagnetic Heisenberg model on a $L\times L$ square lattice with PBC.  SSE  estimates and exact results  are also shown for comparison. Numerical values are given in units of $J_1$.}
\begin{ruledtabular}
\begin{tabular}{lccr}
$L$& This work  & SSE\footnote{Ref. \onlinecite{sand}} & Exact\footnote{Ref. \onlinecite{exact}} \\
\hline
4& -0.7016(1)&-0.701777(7) &-0.7018\\
6 &-0.6785(2)  &-0.678873(4)&-0.6789 \\
8 & -0.6724(3) & -0.673487(4) &- \\
10 & -0.6699(3) & -0.671549(4)&-\\
\end{tabular}
\end{ruledtabular}
\end{table}

The energy per site is a monotonic increasing function of the lattice size. The simple formula\cite{man}
\begin{equation}
E(N)=E_{\infty}-\beta\frac{c}{N^{\frac{3}{2}}}
\end{equation}
where $\beta=1.4377$,\cite{beta} gives an extrapolated energy per site $E_\infty=-0.6683(3)$, which is in good agreement with the most accurate QMC result:\cite{sand} $E_\infty^{SSE}=0.669437(5)$ and lower than other extrapolated values obtained by purely variational methods. For example, on the basis of a Jastrow wave function, Trivedi and Ceperley found $E_\infty^{JST} \simeq -0.6590$.\cite{cep90}

Calculations carried on with open boundary conditions (OBC) yield GS energies lower than those obtained with the general PEPS, or SBS method. For example, we get $E=-0.6258(1)$ for $L=10$ while the PEPS result is $E^{PEPS}=-0.62515$ and the SBS one   $E^{SBS}\sim-0.6225$.
 \begin{table}[b]
\caption{\label{corr} Spin-spin correlation function (computed at the maximum distance on the lattice) of the antiferromagnetic Heisenberg model on a $L\times L$ square lattice with PBC.  SSE  estimates (adjusted for different factors in the definition)   are also shown for comparison. Numerical values are given in units of $J_1$.}
\begin{ruledtabular}
\begin{tabular}{lcr}
$L$& This work  & SSE\footnote{Ref. \onlinecite{sand}} \\
\hline
4&0.1807(4)&0.17962(1)\\
6 &0.1550(4)  &0.152568(9)\\
8 &0.1432(4) &0.13760(1)\\
10 &0.1352(5) &0.12855(2) \\
\end{tabular}
\end{ruledtabular}
\end{table}

Estimates of the spin-spin correlation function computed at the maximum distance on the lattice according to the formula $Corr(L/2,L/2)=\langle \mathbf{S_r} \cdot \mathbf{S_{r'}} \rangle$ where $\mathbf{r}-\mathbf{r'}=(L/2,L/2)$ are shown in Tab \ref{corr}. We find an extrapolated value of the staggered magnetization defined by $M^2(L)=Corr(L/2,L/2)=M^2_\infty+b/L$ of 0.324(1) which is in reasonable agreement with $M^{SSE}_\infty=0.3070(3)$ reported in Ref \onlinecite{sand} and with other estimates.\cite{man} It has to be mentioned  that the discrepancy between our result and the SSE one might be due to a violation of the so-called ``area law".\cite{area}
\begin{table}[t]
\caption{\label{tJ1J2pbc} GS energy per site (in units of $J_1$) for the $J_1-J_2$ model on a square lattice (with PBC) comprising 36 sites.  Exact  results  are also shown for comparison.}
\begin{ruledtabular}
\begin{tabular}{lcr}
$J_2/J_ 1$&This work&Exact\footnote{Ref. \onlinecite{exact}}\\
\hline
0.0&-0.6785(2)&-0.6789\\  
0.1&-0.6377(1)&-0.6381\\  
0.2&-0.5985(1)&-0.5990\\ 
0.3&-0.5616(1)&-0.5625\\  
0.4&-0.5277(1)&-0.5297\\  
0.5&-0.4985(2)&-0.5038\\   
0.6&-0.4860(2)&-0.4932\\ 
0.7&-0.5255(1)&-0.5300\\ 
0.8&-0.5843(1)&-0.5865\\ 
0.9&-0.6453(1)&-0.6491\\ 
1.0&-0.7091(1)&-0.7144\\ 
\end{tabular}
\end{ruledtabular}
\end{table}
\begin{figure}[b]
\centerline{\includegraphics[scale=0.62]{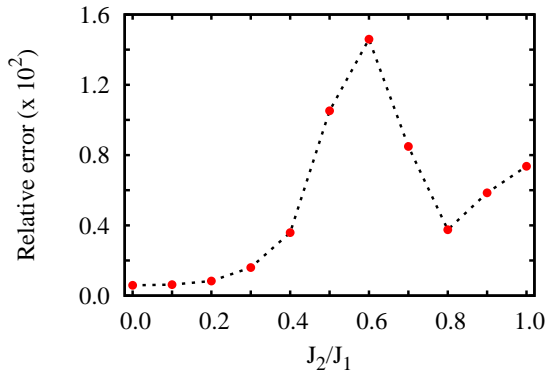}}
\caption{(color online)  Error (relative to the exact result) of the GS energy of the $J_1-J_2$ model computed with the method illustrated in this work. The  square lattice comprises $N=36$ spins; PBC are assumed.  The dashed line is only a  guide to the eye.}
\label{fig:reJ1J2}
\end{figure}

\begin{table}
\caption{\label{tab:PEPS} GS energy per site (in units of $J_1$) for the $J_1-J_2$ model on a square lattice (with OBC) comprising $64$ sites. PEPS results are also shown for comparison.}
\begin{ruledtabular}
\begin{tabular}{lcr}
$J_2/J_ 1$&This work&PEPS\footnote{data provided by V. Murg}\\
\hline
0.0&-0.61567(8)&-0.61506\\
0.1&-0.5865(1)&-0.5845\\
0.2&-0.5571(1)&-0.5555\\
0.3&-0.5290(2)&-0.5274\\
0.4&-0.5028(1)&-0.5016\\
0.5&-0.4781(1) &-0.4779\\
0.6&-0.4523(2)&-0.4508\\
0.7&-0.4553(1)&-0.4541 \\
0.8&-0.4928(2) &-0.4906\\
0.9&-0.5371(1)&-0.5344\\
1.0&-0.5800(1)&-0.5792\\
\end{tabular}
\end{ruledtabular}
\end{table}

By including the next-nearest-neighbor interaction ($J_2 > 0$) in the $J_1-J_2$ Hamiltonian the system becomes frustrated and is believed to undergo a phase transition for $J_2 \simeq 0.6$.
This model cannot be simulated by QMC due to the sign problem  (arising in turn from the underlying Fermi statistics). Our method, instead,  is purely variational and  can be applied without the occurrence of any sign instability.  With the aim of comparing our  estimates with  exact results (available only for  $N \leq 36$), we computed the GS energy, as a function of $J_2/J_1$, of the $J_1-J_2$ model on a $6 \times 6$ square lattice. Numerical values are shown  in Tab. \ref{tJ1J2pbc}; the error of our estimates relative to the exact result is shown in Fig. \ref{fig:reJ1J2}. This quantity never exceeds $1.5\times$10$^{-2}$, moreover it has to be mentioned that the GS energies computed by means of EPS compare favorably to those obtained by GFMC with the fixed-node approximation \cite{massimo} and, except in a narrow region of $J_2/J_1 \sim 0.5$, where the the agreement is however remarkable, to variational results obtained with a  BCS-type ansatz.\cite{sor1,unp}

Also in this case, for OBC, we test  our scheme  against the PEPS one. The estimated energies are  lower than those computed with PEPS even  for a $8 \times 8$ lattices (see Tab. \ref{tab:PEPS}), where the PEPS approach performs at its best.

\section{Conclusions and outlook} 

In this work we have presented Entangled-Plaquette States: an ansatz for the GS wave function of quantum many-body systems which allows the accurate estimate of physical observables by means of Variational Monte Carlo. Our approach not only gives accurate results for unfrustrated systems (where other methods are in principle ``exact" ) but, most importantly, applied to systems for which QMC simulations suffer from a sign problem, yields  estimates whose accuracy seems to be not obtainable with different techniques. Therefore our method appears as a very promising avenue to investigate the effects of Fermi statistics (e.g., frustration).  The extension of our computational approach to 3D can be easily implemented (i.e., taking cubic plaquettes) and several possible improvements (e.g., cover the lattice with plaquettes of different shapes and sizes), are being currently investigated.

\section*{Acknowledgments}
We acknowledge discussions with F. Verstraete and M. Wolf, and thank V. Murg for providing us with the PEPS data. This work has been supported by the DFG (FOR 635) and the EU project QUEVADIS, and the Canadian NSERC through the grant G121210893.

\textit{Note added.---} After this paper was accepted  a related approach has been proposed by J. Hitesh {\it et al.}, {arxiv:0907.4646}.

\end{document}